\documentstyle[12pt]{article}
\begin{document}
\nonumber
{\it Invited talk at 15th European Conference
on Few--Body Problems in Physics, Pe\~niscola, Spain, June 5--9, 1995}
\begin{center}
\begin{large}
{\bf Antiprotonic Helium Atoms} \\[0.3cm]
O.~I.~Kartavtsev \\[0.3cm]
\end{large}
Bogoliubov Laboratory of Theoretical Physics \\
Joint Institute for Nuclear Research\\
141980, Dubna, Russia\\
Fax: 7 096 21 65084\\
E-mail: oik@thsun1.jinr.dubna.su\\
\vspace{1cm}
\end{center}

\sloppy

\begin{abstract}

Metastable antiprotonic helium atoms
$^{3,4}\! H\! e\bar pe$
have been  discovered recently in
experiments of the delayed annihilation of antiprotons in helium
media. These exotic atoms survive for an enormous time (about tens
of microseconds) and carry the extremely large total angular
momentum $L\sim 30-40$.
The theoretical treatment of the intrinsic properties
of antiprotonic helium atoms, their formation and
collisions with atoms and molecules is discussed.

\end{abstract}
\newpage
\section{Introduction}

Metastable antiprotonic helium atoms $^{3,4}\! He\bar pe$ have been
discovered recently in experiments on the delayed annihilation of
antiprotons in helium media~[1--3].
A few percents of antiprotons are captured in these exotic systems,
which survive for an enormous time (about tens of microseconds in
comparison with the mean lifetime of antiprotons in media $\sim
10^{-12}s$) and carry the extremely large total angular momentum
$L\sim 30-40$.  Analogous long-lived systems were observed in
experiments with negative kaons~[4] and pions~[5].
This new class of three-body systems was first predicted in
{}~[6] and theoretically studied in [7]. As far as
the experimental data concern mainly the antiprotonic helium atom,
only this system is considered in this report. One can mention that
described theoretical considerations can be easily applied also to
other hadronic helium atoms.

The observation of the resonant laser-induced
annihilation~[8,9] has initiated thorough
investigations of these unusual systems. Initial populations,
level lifetimes and very precise values of the transition energies
(with the relative accuracy $< 10^{-5}$) have been obtained in these
experiments.

The antiprotonic helium can be considered as a usual helium atom
with one electron replaced by an antiproton. A large angular
momentum $L\sim~(M/m_e)^{1/2}$ ($M - \bar p$--nucleus reduced mass)
provides that  the antiproton-nucleus and
electron-nucleus distances are approximately equal. Besides,
this system can be considered as exotic diatomic molecule, where one
nucleus has charge $-1$.

Main reasons for the enormous
lifetime of antiprotonic helium atoms are as follows.  The
annihilation of an antiproton is inhibited due to the extremely large
total angular momentum  $L \sim 35$.  The Auger decay is
inhibited due to sufficiently large values $( l_e\ge 4)$ of the
angular momentum of an outgoing electron.  Usual mechanism of the
de-excitation by the Stark mixing is not appropriate to the
three-body system due to the lack of degeneracy.  The collisional
de-excitation by surrounding $He$ atoms is suppressed due to the
screening of antiproton by the electron in an antiprotonic helium.
The only remaining de-excitation mechanism is multistep radiative
transitions, whose rates are of order  $\mu s^{-1}$.

For clear understanding of the antiproton fate in the helium medium
one needs to study the slowing down and capture of antiprotons,
intrinsic properties of antiprotonic helium atoms, their decay
processes and interactions with surrounding atoms and molecules. The
purpose of this report is to present the results of existing
theoretical calculations on antiprotonic helium atoms and related
topics with the reference to the modern experimental situation.

\section{Energy levels}

Main problems in the calculation of energy levels arise
from very large values of the total angular momentum $L \sim
30-40$ of antiprotonic helium atoms and instability of these
systems.  However, very small decay rates allow one to regard in
calculations states of these systems as true bound states.

The first calculations of energy levels were made in the pioneering
Russel papers~[7]. Using a simple two-parameter variational
wave function, energy levels were calculated with the relative
accuracy about 3 percents.
The Born-Oppenheimer approximation will be quite reliable
in solving this problem
due to very small mass ratios $m_e/m_{\bar
p}, m_e/m_{He}$
and the next step has been made in the framework
of this approach~[10]. Only ground states were considered in
this paper and more systematic study of antiprotonic helium
eigen-energies within the same approach was carried out in
papers~[11,12] after the discovery of the delayed
antiproton annihilation~[1].  The relative accuracy of
eigen-energies obtained in these calculations can be estimated as
$10^{-4}$.

The configuration interaction calculation~[13] provides a
comparable accuracy of eigen-energies without the reference to the
small mass ratio.  Systematic variational calculations~[14],
with simple trial functions of the form \begin{equation}
\chi_{nml\lambda i}^{LM}({\bf r, r}_1)={\cal Y}_{l\lambda}^{LM}({\bf\hat r,
\hat r}_1) r^{l+i}r_1^\lambda exp(-a_nr-b_mr_1),
\label{eq:trial}
\end{equation}
where $\bf{r, r}_1$ are the electron-nucleus and
antiproton-nucleus
radius-vectors, also provide accuracy of the same order of
magnitude. Very recently, V.~I.~Korobov has obtained more precise
spectra of an antiprotonic helium, using the correlated trial
functions in variational calculations~[15].

By using the method of laser-induced resonant annihilation, two
transition wavelengths $597.259\pm 0.002nm$~[8] and
$470.724\pm 0.002nm$~[9] were measured in the ${^4}H\! e{\bar
p}e$ system and assigned to the transitions $(36,4\rightarrow 35,4)$
and $(35,3\rightarrow 34,3)$, respectively.  The most important for
proving the existence of antiprotonic helium atoms is to compare
experimental and theoretical transition wavelengths
given in Table~1. Here and below in this report the
$L, N$ notation of states, where $L$ is the total angular momentum
and $N$ enumerates the states of the same $L$ value, will be used.\\

 Table~1. Experimental and theoretical transition wavelengths (nm)
 of the $^{4}\! H\! e\bar pe$ system.\\
\begin{tabular}{cccccc}
\hline\hline
Assignment & Experim. & [16] &
[11] & [14] & [15] \\
\hline
$35,4 \rightarrow 34,4 $ & 597.259 &
597.341 & 598.010 & 597.544 & 597.229\\
\hline
$34,3\rightarrow 33,3$ & 470.724 & 470.594 &
 471.351 & 470.871 & 470.705\\
\hline   \hline
\end{tabular}\\

While the agreement of experimental and theoretical transition
wavelengths approves the formation of antiprotonic helium atoms
unambiguously, it is worthwhile to mention a possible ambiguity in
the assignment of the 470.724nm transition. In fact, the calculated
wavelength of the transition $(35,2\to~34,2)$ is also very
close to the experimental value.

In conclusion, a more precise description of energy spectra requires
to take into account the effect of relativistic and QED corrections,
spin-dependent interactions and the coupling with continuous spectrum.

\section{Radiative transitions}

Due to large lifetimes
of metastable states against Auger decay and collisional
de-excitation the radiative transitions become most important in
the description of the system. As far as only dipole transitions
are significant, the total angular momentum changes by unity in
each transition. Thus, the system looses the angular momentum and
energy step-by-step and finally reaches the state with large
Auger decay rate. As a result, the total lifetime is determined by
the number of radiative transitions necessary to reach this state and
rates of each transition.

The dipole transition rates were calculated in papers~[11,13,14,17],
whose results are fairly close to each other. It was found that the
most probable are transitions between states of the same $N$, i. e.
the excitation number is approximately conserved in the radiative
transitions. At the same time, the rates of transitions between
states with different $N$ are smaller at least by 1--2 orders of
magnitude.  Thus, radiative cascades in an antiprotonic helium
proceed almost independently along the chains of states of fixed
$N$.\\
Up to now, lifetimes of the states $(L,N)=(35,4)$ and
$(34,3)$ of $^{4}\!  He\bar pe$ system are determined experimentally
by the method of resonant laser--induced
annihilation~[9,18]. Table 2 contains the
experimental decay rates and theoretical radiative transition rates
of these states. The difference of experimental and theoretical
values is not still understood and may be caused by
additional nonradiative decay channels.

 Table~2. Experimental decay rate and theoretical radiative
 transition rates $(10^6 s^{-1})$ for two states of the $^{4}\!
 H\!  e\bar pe$ system.\\

 \begin{tabular}{ccccc} \hline\hline $L,N$ & Experiment &
[17] & [11] & [14] \\ \hline
 $ 35,4 $ & 0.72 & 0.614 & 0.619 & 0.597 \\ \hline
 $ 34,3 $ & 1.18 & 0.734 & 0.754 & 0.713 \\
 \hline  \hline
 \end{tabular}\\

One can conclude that the existing calculations of radiative
transitions are reasonable accuracy and a detailed
consideration of other decay processes is needed.

\section{Auger decay}

The main feature of the Auger decay rates of antiprotonic helium
atoms is their essential dependence on the multipolarity, i. e.
the angular momentum of the outgoing electron $l_e$. This feature was
supposed already in~[6] and any simple estimate gives the Auger
transition rate of order $10^5s^{-1}$ for the multipolarity
$l_e=4$ and $10^8s^{-1}$ for $l_e=3$.  This
important result was initially obtained in [7].
Bearing in mind the radiative transition rates of order $(10^6
s^{-1})$, one can consider antiprotonic helium states to be
metastable due to the multipolarity of the Auger decay $l_e\ge 4$ .
Calculation of eigen-energies unambiguously determines
multipolarities of the Auger decay and it was performed firstly in
[7] and more exactly in subsequent papers [10,11,13,14].

Up to now the only progress in
this direction is due to calculations of the Auger decay rates of
the $^{4}\! He\bar pe$ system~[16,17].  In these
calculations, an increase of the multipolarity from $l_e= 3$
to $l_e= 4$ as a rule gives a decrease of the Auger rate by
three orders of magnitude. Besides, for some states the calculated
Auger rates are much smaller than the typical values prescribed by
the multipolarity.  This interesting peculiarity is a result of the
interference of different parts of the initial and
final state wave functions.

It is worthwhile to mention that these results are based on the
approximation of the continuous spectrum final state wave function.
Namely, the effective two--body model was used to describe the
interaction of the outgoing electron and the hydrogenlike $\bar
pHe^{++}$ system.

As it was mentioned in the preceding section, the experimental data
call for the intensive theoretical study of decay processes and
precise few--body calculations are desirable for understanding the
cascade processes in an antiprotonic helium.

\section{Energy levels' splitting and relativistic corrections}

As it was discussed above, the precise measurement
of transition energies of antiprotonic helium atoms in recent
experiments on the laser-induced resonant annihilation invokes the
calculations of comparable accuracy.  For this reason,
relativistic corrections of order $\alpha ^2$ (in a.u.)  to
the pure Coulomb interaction should be taken into account in the
precise calculations of energy spectra of antiprotonic helium atoms.
Since the precision of experiments can be improved significantly,
also QED corrections to energies of higher orders on $\alpha $ will
be calculated. Precise calculations and measurements of the
energy spectra can be also used for the determining the
antiproton properties, e.g.  the magnetic moment. Experimentally,
the most simple is the measurement of the energy-level splitting
due to the spin-dependent part of the relativistic interaction.

Recently, the energy-level splitting of antiprotonic helium atoms
has been calculated with the wave functions~[14]. The details
of this calculation and analysis of the level structure will be
presented in~[19]. Short results of this calculation
are given below.

The splitting of levels arise due to the spin-dependent part of the
Breit interaction in each pair of particles in antiprotonic helium
atoms.  Since the magnetic moments and velocities of particles
scale inversely proportional to the particle mass and very small
ratios $m_e/m_{H\! e}, m_e/m_{\bar p}$, the largest contribution to
the energy splitting comes from the interaction with the electron
spin.  This part of relativistic interaction can be written as
follows:

\begin{equation}
H_{s}=\alpha ^2\! \sum_{i=\bar p,H\! e}\, \frac{Z_i}{r_{ei}^3}\, {\bf
s}_e\cdot {\bf r}_{ei}\times ({\textstyle\frac{1}{2}}{\bf v}_{e}-{\bf
v}_i) ,
\end{equation}
where $\alpha $ is the fine structure constant,
${\bf r}_{i}, {\bf v}_{i}, {\bf s}_{i}, {\bf Z}_{i}$
are the radius-vector, velocity, spin and charge of particle $i$,
${\bf r}_{ei}={\bf r}_{e}-{\bf r}_{i}$.\\
 This interaction conserves the sum ${\bf j}={\bf L}+{\bf s_e}$ of
 the total angular momentum $\bf L$ and electron spin ${\bf s}_e$ and
splits each level into two sublevels, corresponding to the
eigenvalues $j=L\pm 1/2$.

Splitting values, defined as $\Delta E_L=E(j=L+1/2)-E(j=L-1/2)$,
have been calculated in the first order
of perturbation theory over $H_{s}$ by using variational
nonrelativistic wave functions~[14].
These values for a number of states of $^{4}\! He\bar pe$
systems in the range $32\leq L\leq 37$ are presented in Table 3.\\

Table~3. Splitting values $\Delta E_L (10^{-4} eV)$ of the $^{4}\!
H\! e\bar pe$ system.

\begin{tabular}{cccccc} \hline\hline
$N \backslash  L$ & 33 & 34 & 35 & 36 & 37 \\ \hline
$ 1 $ & -.311 & -.313 & -.311 & -.305 & -.298 \\ \hline
$ 2 $ & -.296 & -.294 & -.289 & -.282 & -.272 \\ \hline
$ 3 $ & -.278 & -.273 & -.266 & -.256 & -.245 \\ \hline
$ 4 $ & -.256 & -.255 & -.246 & -.234 & -.223 \\ \hline
$ 5 $ & -.277 & -.252 & -.245 & -.228 & -.226 \\ \hline   \hline
\end{tabular} \\

Experimentally, only the difference in splitting of two levels can
be observed. Since the dependence of calculated splitting values on
$L$ is rather slow, it is not possible to resolve such a small
difference in splittings for the favoured transitions, i. e.
transitions between states of the same $N$. For this reason, the
experimental proposal for near future~[20] is aimed at searching
the splitting in the unfavoured  transitions $(L,N) \to (L-1,N+2)$.

The part of interaction depending on heavy particle
spins removes the remaining degeneracy and split each $j=L\pm 1/2$
sublevel further into two or four levels for $^{4}\! H\! e\bar
pe$ and $^{3}\! H\! e\bar pe$ systems, respectively. Values of
this secondary splitting are much smaller in comparison with the
initial splitting.

\section{Formation probability and initial populations }

The calculation of the formation probability and initial populations
of antiprotonic helium atoms is important for the understanding of
cascade processes and the interpretation of experimental data.
Antiprotonic helium atoms are produced in the reaction of the
antiproton capture by a helium atom  with an electron emission. As
antiprotons mainly loose energy in inelastic collisions, one can
assume that their initial kinetic energy does not exceed the
ionization energy of helium atoms.  Calculations of the antiproton
capture and slowing down in helium~[21], made in the framework of
semiclassical model, confirm this assumption.

The formation probability of $0.22$ for one
stopped antiproton, calculated in this paper, is far from
the experimental value $0.03-0.04$.  The possible
reason for this difference is due to the model used in the
calculation. One can mention that calculated energies of the
considerable part of final states are smaller in comparison with
results of precise calculations.

The important qualitative result, obtained in this calculation, is
a strict connection on the initial kinetic energy of an antiproton
and the total angular momentum of the formed antiprotonic helium
atom. As a consequence of this connection the formation of
antiprotonic helium atoms with the fixed angular momentum is due to a
capture of an antiproton with the fixed energy.

Another approach, used quasi-classical conceptions, have been applied
to the calculation of initial populations of antiprotonic
helium atoms~[22]. Lower bound for the populations of antiprotonic
helium atoms have been obtained. Thus, only states of the large
enough $L$ and $N$ values can be formed. Full formation probability in
this calculation is about 0.3 per one stopped antiproton and also
differs from the experimental value.

Up to now, the experimental information on populations of states for
$N=3, 4$ near the metastability boundary have been
obtained~[9,18].  One can mention the necessity of the reliable
calculations of the formation probability and initial populations by
using few-body methods for the precise solution of this problem.

\section{Interaction with helium medium}

The interaction of antiprotonic helium atoms with the surrounding
medium plays a crucial role in its survival. The angular momentum and
energy transfer is hindered due to the neutrality of antiprotonic
helium, screening of the antiproton by the electron and large
excitation energy of the ordinary helium atom. This qualitative
arguments are supported by the experiment~[3,23],
where only a small, although significant, density dependence of the
delayed annihilation time spectra have been found.

At the same time an average antiproton lifetime, transition
lifetimes and a fraction of the so-called fast component in the
delayed annihilation time spectrum depend on density. An average
antiproton lifetime depends also on the phase of the helium medium
and its value in solid helium is 20 percent shorter than in liquid
helium of the same density. One more problem is the observed isotope
dependence of the decay rate of the fast component~[23].  One
should obtain the quantitative description of the existing
density and phase dependence.

Only calculations of the ordinary helium atom -- antiprotonic helium
interaction were obtained~[22] in the framework of
multi-channel approach. The quenching of the delayed annihilation
due to processes $$ \bar pHe^+(L_iN_i)+ He \to \bar pHe^+(L_fN_f)+ He $$ in
collisions with surrounding helium atoms has been considered. Main
conclusions are as follows:\\
The states with $L+N \leq 40 $ are stable against collisional
quenching at all densities up to the density of liquid helium.
 The states with $L+N=41$ are intermediate: they are destroyed by
 collisions at high densities.
The states with $ L+N \geq 42 $ are fully destroyed at
$ p \geq 1 bar $, and can survive at very low densities
($p \sim 1 mbar $).

One can conclude that the influence of medium on antiprotonic helium
atoms, including the external Auger effect, collisional de-excitation,
collisional broadening of radiation spectral lines and solid phase
effect are still awaiting for their investigation.

\section{Quenching by impurities}

The  effect of impurities on the delayed annihilation was
discovered in experiment~[2,3,24]. The admixtures of noble
gases ($Ne, Ar, Kr$) of order 10 percent shorten the average
$\bar p$ lifetime
slightly. The effect of molecular gases ($H_2, N_2, O_2$) is
noticeable at impurity concentrations as low as $20 ppm$. For the
most investigated case of the molecular hydrogen the following
estimate for the quenching cross-section $\sigma _q$ was found:
$\sigma _qv=6\cdot 10^{-11}cm^3/s$. Thus the quenching cross-section
is close to the geometrical cross-section, i. e. metastable states
are destroyed in a single collision with a hydrogen molecule.

The qualitative understanding of these facts is based on the easy
possibility to transfer the angular momentum to a diatomic molecule.
On the contrary, such a possibility does not exist for  noble gases
due to large excitation energies.

Up to now there are no calculations of the interaction of
antiprotonic helium atoms and another atoms or molecules.
At the same time, delayed annihilation time
spectra reveal some peculiarities in the dependence on the impurity
concentration. As a result, the theoretical study is
necessary to provide the quantitative description of observed
quenching phenomena.

\section{Other exotic systems and reactions}

The discovery of antiprotonic helium atoms caused searching of
other long-lived systems with an antiproton. The most probable
candidates are systems consisted of an antiproton captured by the noble
gas atom. Another possibility is the antiprotonic lithium and
energies of this system were calculated in~[10].  However, no
delayed annihilation was detected experimentally in gaseous $Ne$~[3],
metallic $Li$ and $LiH$~[23].  Calculations of formation and
quenching rates in these media are necessary to clear up the
situation.

Besides of neutral antiprotonic helium atoms, $\bar pH\! e$
systems of charge $-1$ can be formed~[25] due to the attractive
polarization $\bar p$ -- $H\! e$ potential. If these systems survive
for sufficiently long time, they can contribute to the kinetics of
antiprotons in media.

New exotic systems can be produced in reactions of long-lived
antiprotonic helium atoms with other atoms and molecules. In
particular, the reaction $$\bar pHe^+ + Ps \to \bar H + He$$ was
proposed~[26] for the antihydrogen production. One can assume also
the possibility of exotic molecules, which include as a part the
antiprotonic helium atom.

\section{Conclusion}

Nowadays new three-body systems are discovered and intensively
investigated both theoretically and experimentally. The description
of these exotic systems meet some difficulties due to
metastability and the extremely large total angular momentum $L\sim
30-40$.

There is the principal possibility to study antiproton properties
by using measurements of antiprotonic helium atoms.

A number of problems arise in the theoretical treatment of
intrinsic properties of antiprotonic helium atoms, their formation
and collisions with atoms and molecules. These problems represent a
wide area for the application of few--body methods of calculations.

{\it Acknowledgement.}
Fruitful discussions with V.~Belyaev, J.~Eades, R.~Hayano,
G.~Korenman, V.~Korobov, K.~Ohtsuki, I.~Shimamura, E.~Widmann and
T.~Yamazaki were of great importance in the preparation of this
report. The author is grateful to the University of Valencia and
Russian Academy of Sciences for financial support in the
participation in Few--Body 15.

\end{document}